\documentclass[12pt]{article}
\usepackage{cite,psfig}

\def\be{\begin{equation}}
\def\ee{\end{equation}}
\def\xbj{x_{\mbox{\scriptsize Bj}}}
\def\bea{\begin{eqnarray}}
\def\eea{\end{eqnarray}}
\newcommand{\ba}{\begin{eqnarray}}
\newcommand{\ea}{\end{eqnarray}}
\newcommand{\mst}[2]{\mbox{\raisebox{-1mm}{$\,\stackrel{#1}{\scriptstyle 
#2}\,$}}}
\renewcommand{\theequation}{\arabic{section}.\arabic{equation}}

\begin{document}

\vspace*{.4cm}
\noindent
{\large HD-THEP-99-40\hfill September 1999}
\vspace*{2.3cm}

\begin{center}
\renewcommand{\thefootnote}{\fnsymbol{footnote}}
{\Large\bf Structure functions at small $\xbj$ \\[.3cm]
in a Euclidean field theory approach\footnote[1]{Supported 
by BMBF, contract No. 05 HT9VHA3.} }\\[2.5cm]
\renewcommand{\thefootnote}{\arabic{footnote}}
{\large A. Hebecker, E. Meggiolaro and O. Nachtmann}\\[.5cm]
{\it Institut f\"ur Theoretische Physik der Universit\"at Heidelberg\\
Philosophenweg 16, D-69120 Heidelberg, Germany}\\[.5cm]
{\small a.hebecker@thphys.uni-heidelberg.de, 
e.meggiolaro@thphys.uni-heidelberg.de, o.nachtmann@thphys.uni-heidelberg.de}
\\[2.1cm]

{\bf Abstract}\end{center}
\noindent
The small-$\xbj$ limit of deep inelastic scattering is related to the 
high-energy limit of the forward Compton amplitude in a familiar way. We 
show that the analytic continuation of this amplitude in the energy 
variable is calculable from a matrix element in Euclidean field theory. 
This matrix element can be written as a Euclidean functional integral in an 
effective field theory. Its effective Lagrangian has a simple expression in 
terms of the original Lagrangian. The functional integral expression 
obtained can, at least in principle, be evaluated using genuinely 
non-perturbative methods, e.g., on the lattice. Thus, a fundamentally new 
approach to the long-standing problem of structure functions at very small 
$\xbj$ seems possible. We give arguments that the limit $\xbj\to 0$ 
corresponds to a critical point of the effective field theory where the 
correlation length becomes infinite in one direction. 
\thispagestyle{empty}
\newpage

\section{Introduction}
The behaviour of the structure functions of deep inelastic lepton-nucleon
scattering (DIS) at small values of the Bjorken variable $\xbj$ has been
a topic of intense experimental and theoretical research in recent years.
The strong rise of the structure functions for $\xbj\to 0$ observed at
HERA~\cite{hera} opened the way for a more detailed experimental study
of the small-$\xbj$ limit of QCD. Today, many theoretical models try to
explain the behaviour of the structure functions in this limit.

In a very pragmatic perturbative approach, the well-known DGLAP dynamics
of the $Q^2$ evolution of structure functions~\cite{dglap} is taken
seriously in the small-$\xbj$ region (cf., for example, the parametrizations
of~\cite{pdfs}). Given a sufficiently small $Q^2$ starting scale of the 
perturbative evolution, the small-$\xbj$ rise of structure functions can 
even be described on the basis of valence-like parton distributions 
(see~\cite{grv} and refs. therein). In this case, all of the small-$\xbj$ 
rise is produced by perturbative QCD. 

If, in a more modest approach, a constant behaviour or a soft growth for 
$\xbj\to 0$ is ascribed to the input distributions, the DGLAP evolution 
allows for a good fit to the data even in the case of a larger $Q^2$ 
starting scale. This observation underlies successful analyses based on the 
idea of double asymptotic scaling~\cite{das}. It is also at the heart of a 
recent combined analysis of diffractive and inclusive structure 
functions~\cite{bgh}, where a universal logarithmic growth of the input 
distributions was assumed. 

An obvious problem of the above, purely DGLAP-based approaches is the
presence of large $\ln(1/\xbj)$ corrections, which can spoil the usefulness 
of the perturbation series as an asymptotic expansion. Using the BFKL 
resummation method~\cite{bfkl} (see~\cite{lipr} for reviews and further 
references), all leading $\ln(1/\xbj)$ contributions can be included in 
the structure function analysis (see, e.g.,~\cite{f2a}). It is, at present, 
not completely clear whether the predicted very strong power-like growth is 
borne out by HERA data. Also, no final conclusions can yet be made 
concerning the effect of the recently obtained next-to-leading order 
results in the BFKL framework (see~\cite{nlo} and refs. therein) on 
structure function analyses. Different methods of dealing with the 
apparent large size of the corrections have been suggested~\cite{mnlo}. 

Unfortunately, because of the problem of infrared diffusion~\cite{mue}, it 
is not clear that the perturbative resummation of $\ln(1/\xbj)$ terms at 
any given order in $\alpha_s$ can reveal the asymptotic behaviour. One 
possibility to go beyond the above resummation schemes is the construction 
of a high-energy effective action for reggeons (see~\cite{lip} and refs. 
therein). Another possibility is the derivation of a small-$\xbj$ evolution 
equation in the framework of the semiclassical approach. This method goes 
back to the treatment of hadron-hadron scattering in the functional 
integral approach using the eikonal approximation introduced in~\cite{eik}. 
For recent results concerning the energy dependence of cross sections in 
this framework see, e.g.,~\cite{sc}. 

A very different point of view is advertised in \cite{dl}, where, in
addition to the familiar soft pomeron, a phenomenological hard pomeron is
introduced to describe small-$\xbj$ structure functions. A further
analysis~\cite{cdl}, motivated by the successful phenomenology
of~\cite{dl}, concludes that the appearance of new small-$\xbj$
singularities due to $Q^2$ evolution is inconsistent with the analyticity
principle governing Regge theory and underlying the approach of~\cite{dl}.
Note also the successful analysis of various experimental cross sections 
reported in~\cite{rue}, which combines ideas of~\cite{eik} and~\cite{dl}. 

In spite of the large amount of work invested and the good description of
data achieved in most of the above approaches, our fundamental understanding
of the small-$\xbj$ asymptotics in QCD is still not satisfactory. Therefore,
we propose yet another method to achieve this goal. In this article, we
will derive a connection between the small-$\xbj$ behaviour of the
structure functions of DIS and the behaviour of correlation functions
in a certain effective field theory, which we consider both in Minkowski
and Euclidean space-time. The analytic continuation from Minkowski to
Euclidean space-time to be discussed below is similar in spirit to the one
described in \cite{meg} for the high-energy hadron-hadron scattering
amplitudes. However, the details are quite different and we will comment on
the connection between the two approaches.

For simplicity, we study in this article a scalar model field theory where
scalar currents replace the electromagnetic currents of real life, i.e.,
QCD. However, it will become clear that our methods allow a straightforward
generalization to QCD.

Our article is organized as follows. In Sect.~\ref{camp} we define the 
model and discuss the properties of our scalar analogue of the virtual 
Compton amplitude. In Sect.~\ref{mieu} we introduce a class of effective 
Hamiltonians depending on a parameter $r$. We also make a rotation from 
Minkowski to Euclidean space-time and show that the high-$Q^2$, $\xbj\to 0$ 
limit corresponds to $r\to 0$. In Sect.~\ref{fi} the amplitude is written 
as a functional integral with $r$-dependent action. In Sect.~\ref{ff} the 
formalism is applied to a simple model of free fields. We show that -- at 
least in the free-field limit -- we get a diverging correlation length in 
our effective theory for $r\to 0$, i.e., a critical phenomenon. Finally, 
two alternative continuations from Minkowski to Euclidean space-time are 
outlined in Sect.~\ref{alt}. Section~\ref{conc} contains our conclusions.

\section{The model and the scalar analogue of the virtual Compton
amplitude}\label{camp}
\setcounter{equation}{0}

We consider a model with one real scalar field $\phi(x)$ and the classical 
Lagrangian 
\be
{\cal L}(x)=\frac{1}{2}\,\partial_\mu\phi(x)\,\partial^\mu\phi(x)-\frac{m^2}
{2}\,\phi(x)^2-\frac{\lambda}{4!}\,\phi(x)^4\,,\label{lag}
\ee
where $m$ and $\lambda$ are the unrenormalized mass and coupling parameter, 
respectively. The momentum canonically conjugate to $\phi(x)$ is
\be
\pi(x)=\frac{\partial{\cal L}}{\partial \dot{\phi}(x)}=\dot{\phi}(x)\,.
\ee
The Hamiltonian and the third component $P^3$ of the momentum operator read 
\be
H=\int_{x^0=\mbox{\scriptsize const.}}d^3\vec{x}\left\{\frac{1}{2}\Pi(x)^2
+\frac{1}{2}\Big(\nabla\Phi(x)\Big)^2+\frac{m^2}{2}\Phi(x)^2+\frac{\lambda}
{4!}\Phi(x)^4\right\}\,,\label{hop}
\ee
\be
P^3=-\frac{1}{2}\int_{x^0=\mbox{\scriptsize const.}}d^3\vec{x}\,\left(\Pi(x)
\frac{\partial\Phi(x)}{\partial x^3}+\frac{\partial\Phi(x)}{\partial x^3}
\Pi(x)\right)\,.\label{pop}
\ee
In Eqs.~(\ref{hop}) and (\ref{pop}) no vacuum expectation values are 
subtracted. Note that we use lower- and upper-case letters for the 
classical field variables and the corresponding operators respectively. 

\begin{figure}[ht]
\begin{center}
\vspace*{.2cm}
\parbox[b]{5cm}{\psfig{width=5cm,file=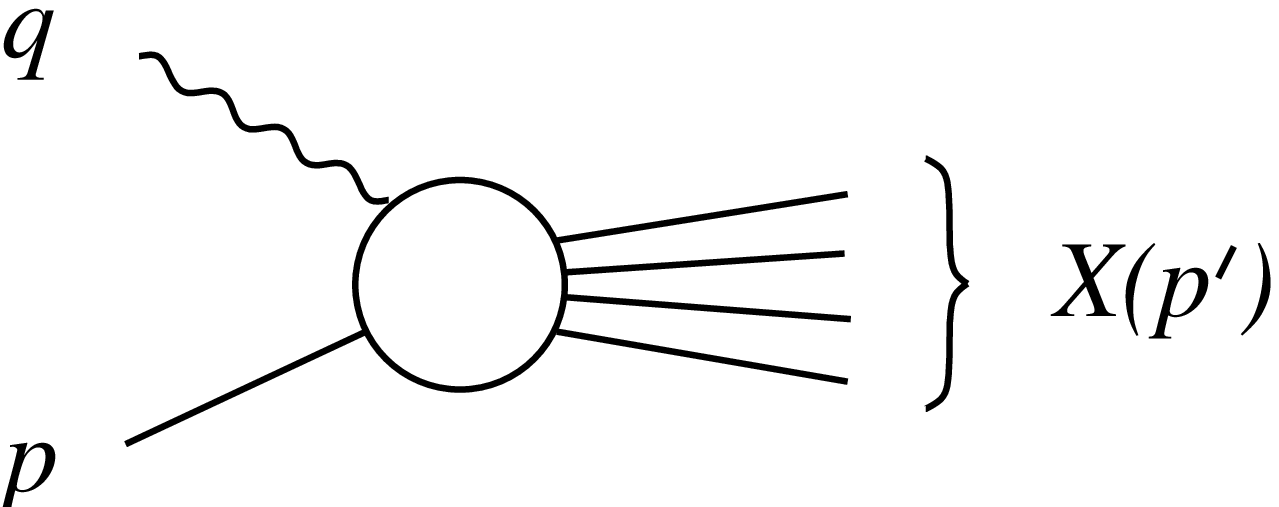}}\\
\end{center}
\refstepcounter{figure}\label{dis}
{\bf Figure \ref{dis}:} Diagram for the absorption of a scalar photon on
the scalar proton.
\end{figure}

We assume that the theory has physical states with reasonable
spectrum\footnote{
Of course we know that the theory defined by Eq.~(\ref{lag}) is trivial if
the cutoff goes to infinity in four dimensions. The purist may always think of
a theory with a finite cutoff.}
etc. We will call `proton' the particle of lowest mass $M$. We introduce
now an `electromagnetic' coupling of our field to a scalar `photon' with
field $A(x)$:
\be {\cal L}'(x)=-e\,J(x)\,A(x)\qquad\mbox{with}\qquad
J(x)=\Phi(x)^2\,.\label{je}
\ee
This scalar current is our analogue of the electromagnetic current of QCD.
In the following we will study DIS, i.e., the cross section for the
absorption of the scalar `photon' on the scalar `proton' (Fig.~\ref{dis}).

We define the structure function for this reaction as
\be
W(\nu,Q^2)=\sum_X\,\frac{1}{2}(2\pi)^3\delta^{(4)}(p'-p-q)\,\langle P(p)
|J(0)|X(p')\rangle\,\langle X(p')|J(0)|P(p)\rangle\,,\label{sf}
\ee
where
\be
Q^2=-q^2>0\qquad\mbox{and}\qquad \nu=pq/M>0\,.
\ee
All the notation is standard. The conventions for the metric, normalization
of states etc. follow~\cite{nac}.

The object to study from a theoretical point of view is the virtual Compton
amplitude. To be precise, we will study the amplitude corresponding to the
retarded commutator (see, e.g,~\cite{kal}):
\be
T_r(\nu,Q^2)=\frac{i}{2\pi}\int d^4x\,e^{iqx}\theta(x_0)\,\langle P(p)|
[J(x),J(0)]|P(p)\rangle\,.\label{tr}
\ee
Here and in the remainder of the paper, we always assume that the connected 
part of the relevant matrix element is taken, $\langle\cdots\rangle= 
\langle\cdots\rangle_c$. It is easy to see that
\be
\mbox{Im}\,T_r(\nu+i\varepsilon,Q^2)=\mbox{sgn}(\nu)\,W(|\nu|,Q^2)\label{imt}
\ee
for $\nu$ real and $\varepsilon \to 0+$.

We consider now Eq.~(\ref{tr}) in the proton rest frame, where we can set
\be
p=(M,\vec{0})\qquad,\qquad q=(\nu,\,\vec{e}_3\sqrt{\nu^2+Q^2}\,)\label{pq}
\ee
with $\vec{e}_j\,\,(j=1,2,3)$ the spatial cartesian unit vectors.
Rotational invariance gives
\be
\langle P(p)|J(\vec{x},x^0)J(0)|P(p)\rangle=\langle P(p)|J(\pm|\vec{x}|
\vec{e}_3,x^0)J(0)|P(p)\rangle\,.\label{rot}
\ee
Inserting this in Eq.~(\ref{tr}) and using Eq.~(\ref{pq}) and the fact that 
the commutator in Eq.~(\ref{tr}) vanishes outside the light cone, we find 
\be
T_r(\nu,Q^2)=-\frac{1}{\sqrt{\nu^2+Q^2}}\int\limits^\infty_0dx^0
\int\limits^{x^0}_{-x^0}dx^3x^3\,e^{ix^0\nu-ix^3\sqrt{\nu^2+Q^2}}\,
\langle P(p)[J(x^3\vec{e}_3,x^0),J(0)]|P(p)\rangle\,.\label{trm}
\ee

\begin{figure}[ht]
\begin{center}
\vspace*{.2cm}
\parbox[b]{7cm}{\psfig{width=7cm,file=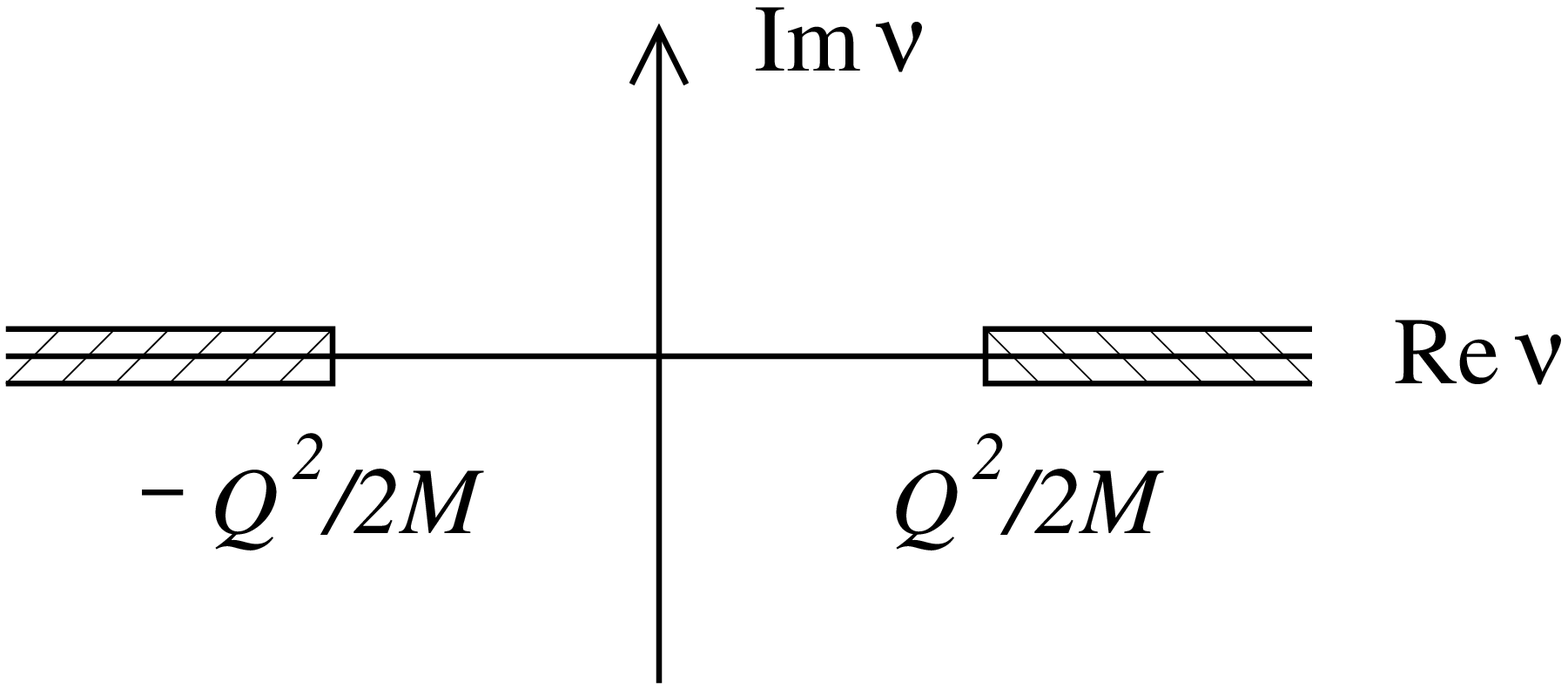}}\\
\end{center}
\refstepcounter{figure}\label{nup}
{\bf Figure \ref{nup}:} 
The analyticity structure of the virtual Compton amplitude $T_r(\nu,Q^2)$ 
in the $\nu$ plane. Singularities can only occur for real $\nu$, $|\nu|\ge 
Q^2/2M$. The integral in Eq.~(\ref{trm}) is a representation of $T_r(\nu, 
Q^2)$ for Im$\,\nu>0$. 
\end{figure}

In Eq.~(\ref{trm}) we have a convenient starting point for the analytic 
continuation of $T_r(\nu,Q^2)$ into the upper half of the complex $\nu$ 
plane, keeping always $Q^2$ fixed. In fact, it is easy to see that the 
integral in Eq.~(\ref{trm}) defines an analytic function for Im$\,\nu>0$ 
(cf.~Appendix~\ref{rac}). Using standard methods, one finds then that 
$T_r(\nu,Q^2)$ can be extended to a function analytic in the whole $\nu$ 
plane except for poles and cuts on the real axis for $|\nu|\ge Q^2/2M$ 
(cf.~Fig.~\ref{nup}). 

We are interested in the behaviour of $T_r(\nu,Q^2)$ for $\nu\to\infty$ on 
the real axis. Instead of studying this directly, we will study $T_r(\nu, 
Q^2)$ on the imaginary axis, i.e., for $\nu=i\eta,$ $\eta\to +\infty$. With 
standard assumptions for the high-momentum behaviour of the amplitudes of 
QFT (cf.~Appendix~\ref{rac}), we can apply the Phragm\'{e}n-Lindel\"of 
theorem (see, e.g.,~\cite{pl}), which ensures that the asymptotic behaviour 
on the real and imaginary axes are related by analytic continuation. Our 
object of study is thus: 
\be
T_r(i\eta,Q^2)=\frac{i}{\sqrt{\eta^2-Q^2}}\int\limits^\infty_0dx^0
\int\limits^{x^0}_{-x^0}dx^3 x^3\,e^{-x^0\eta+x^3\sqrt{\eta^2-Q^2}}\,
\langle P(p)|[J(x^3\vec{e}_3,x^0),J(0)]|P(p)\rangle\,.
\ee
It turns out to be convenient to generalize this amplitude slightly. Let 
$\mu$ with Im$\,\mu\geq0$ be a complex parameter with dimension of mass and 
define: 
\be
T_+(i\eta,Q^2,\mu)=\frac{2i}{\sqrt{\eta^2-Q^2}}\!\int\limits^\infty_0\!dx^0
\!\!\int\limits^{x^0}_{-x^0}\!dx^3x^3\,e^{x^0(i\mu-\eta)+x^3
\sqrt{\eta^2-Q^2}}\,\langle P(p)|J(x^3\vec{e}_3,x^0)J(0)|P(p)\rangle.
\label{tp}
\ee
We have
\be
T_r(i\eta,Q^2)=\mbox{Re}\,T_+(i\eta,Q^2,0)\,,\label{trtp}
\ee
which is obvious for $\eta>Q$ and requires a little thought for $0<\eta\le 
Q$ (cf.~Appendix~\ref{rac}). 

In the next section we will represent $T_+(i\eta,Q^2,\mu)$ for Re$\,\mu<-M$ 
as a correlation function in an effective Euclidean field theory.
The amplitude $T_r(i\eta,Q^2)$ is then obtained from Eq.~(\ref{trtp}) after 
analytic continuation of $T_+$ to $\mu=0$. We will also write
$T_+(i\eta,Q^2,0)$ as a correlation function in the corresponding
effective theory in Minkowski space-time.

\section{From Minkowski to Euclidean space}\label{mieu}
\setcounter{equation}{0}
Let us introduce new coordinates in Eq.~(\ref{tp}):
\be
x^0=\xi\ \cosh\chi\quad,\quad x^3=\xi\ \sinh\chi
\ee
with
\be
0\leq\xi<\infty\quad,\quad-\infty<\chi<\infty\,.
\ee
In this section we always consider $\eta>Q$.
Then we insert a sum over a complete set of states and perform the $\xi$ 
integration: 
\bea
&&\hspace*{-.5cm}
T_+(i\eta,Q^2,\mu)=\sum_X\frac{2i}{\sqrt{\eta^2-Q^2}}\int^\infty_{-\infty}
d\chi\,\sinh\chi\ \int^\infty_0d\xi\,\xi^2\label{tps}
\\
&&\hspace{.5cm}\times e^{-\xi\left[\left(\eta+i({p'}^0-M-\mu)\right)\cosh
\chi-(\sqrt{\eta^2-Q^2}+i{p'}^3)\sinh\chi\right]}\,\langle P(p)|J(0)|X(p')
\rangle\langle X(p')|J(0)|P(p)\rangle\,.\nonumber
\eea
The convergence of the $\xi$ integration in Eq.~(\ref{tps}) is guaranteed 
since
\be
\eta\ \cosh\chi-\sqrt{\eta^2-Q^2}\sinh\chi>0\,,
\ee
and we get
\bea
T_+(i\eta,Q^2,\mu)&=&\sum_X\frac{2i}{\sqrt{\eta^2-Q^2}}\int^\infty_{-\infty}
d\chi\ \sinh \chi\label{tpn}
\\
&&\times 2\left\{\left(\eta+i({p'}^0-M-\mu)\right)\cosh\chi-\left(
\sqrt{\eta^2-Q^2}+i{p'}^3\right)\sinh\chi\right\}^{-3}\nonumber
\\
&&\times\langle P(p)|J(0)|X(p')\rangle\langle 
X(p')|J(0)|P(p)\rangle\,.\nonumber
\eea
Since ${p'}^0\geq |{p'}^3|$, we have also
\be
({p'}^0-M-\mbox{Re}\,\mu)\cosh\chi-{p'}^3\sinh\chi>0
\ee
for Re$\,\mu<-M$. This allows us to write Eq.~(\ref{tpn}) again as an 
integral over a parameter~$\xi$:
\bea
T_+(i\eta,Q^2,\mu)&=&\sum_X\frac{-2}{\sqrt{\eta^2-Q^2)}}
\int^\infty_{-\infty} d\chi\sinh\chi\ \int^\infty_0 d\xi\xi^2\label{tpl}
\\
&&\times\,e^{-\xi\left[({p'}^0-M-\mu-i\eta)\cosh\chi-({p'}^3-i
\sqrt{\eta^2-Q^2})\sinh\chi\right]}\nonumber
\\
&&\times\langle P(p)|J(0)|X(p')\rangle\langle X(p')|J(0)|P(p)\rangle 
\nonumber
\\
\nonumber\\
&=&\frac{-2}{\sqrt{\eta^2-Q^2}}\int^\infty_{-\infty}d\chi\sinh\chi
\int^\infty_0d\xi\xi^2\nonumber
\\
&&\times\,e^{\xi\left[(\mu+i\eta)\cosh\chi-i\sqrt{\eta^2-Q^2}\sinh\chi
\right]}\,{\cal M}_E\,,\nonumber
\eea
where
\be
{\cal M}_E=\langle P(p)|\,e^{\xi(H\cosh\chi-P^3\sinh\chi)}\,J(0)\,
e^{-\xi(H\cosh\chi-P^3\sinh\chi)}\,J(0)|P(p)\rangle\,.
\label{me}
\ee
Effectively we have shifted the $\xi$-integration in Eq.~(\ref{tps})
from the real to the imaginary axis, $\xi\to -i\xi$. In Eq.~(\ref{me})
$H$ and $P^3$ denote the Hamilton operator and the third component of the 
momentum operator of the original theory, Eqs.~(\ref{hop}) and (\ref{pop}), 
respectively.

The matrix element ${\cal M}_E$ can now be interpreted as one of a 
Euclidean field theory. For this we set
\be
y^0=\xi\cosh\chi\qquad,\qquad y^3=\xi\sinh\chi\qquad,\qquad
y_\pm=y^0\pm y^3=\xi e^{\pm\chi}
\ee
and
\be
r=\frac{2y_-}{y_++y_-}=1-\frac{y^3}{y^0}\qquad,\qquad 
\eta_\pm=\eta\pm\sqrt{\eta^2-Q^2}\,.\label{reta}
\ee
We have
\be
y_+\geq0,\quad y_-\geq0\qquad,\qquad 0\leq r\leq2\,.
\ee
Furthermore we define an effective, $r$-dependent Hamiltonian: 
\be
H_{eff}(r)=H-(1-r)P^3\,.\label{heff}
\ee
With this we can write
\be
{\cal M}_E\equiv{\cal M}_E(y_0,r)=\langle P(p)|\,e^{y^0H_{eff}(r)}\,J(0)\,
e^{-y^0H_{eff}(r)}\,J(0)|P(p)\rangle\label{eme}
\ee
and
\bea
T_+(i\eta,Q^2,\mu)&\!\!\!=\!\!\!&
\frac{-1}{2\sqrt{\eta^2-Q^2}}\int^\infty_0 dy_+
\int^\infty_0dy_-(y_+-y_-)\,e^{\frac{1}{2}y_+(\mu+i\eta_-)+\frac{1}{2}
y_-(\mu+i\eta_+)}\,{\cal M}_E(y^0,r)\nonumber
\\
\label{tpme}\\
&\!\!\!=\!\!\!&
\frac{-2}{\sqrt{\eta^2-Q^2}}\int^\infty_0dy^0(y^0)^2\int^2_0dr(1-r)\,
e^{y^0[\mu+i(1-\frac{r}{2})\eta_-+i\frac{r}{2}\eta_+]}\,{\cal M}_E(y^0,r)
\,.\nonumber
\eea
We see from Eq.~(\ref{reta}) that for $\eta\to\infty, \ Q^2$ fixed,
we have $\eta_+\sim 2\eta$, $\eta_-\sim Q^2/(2\eta)$. Then the
oscillating term $\exp(i y^0r\eta_+/2)$ restricts the
$r$-integration to
\be
r\mst{<}{\sim}\frac{1}{\eta y^0}\,.
\ee
Thus, except for the small-$y^0$ range, $0\,{\scriptstyle \leq}\,y^0\mst{<}
{\sim}1/ \eta$, the behaviour of the matrix element ${\cal M}_E(y_0,r)$ for 
$r\to 0$ will be essential. Going from $r=1$ to $r=0$ in Eq.~(\ref{heff}) 
corresponds, of course, to going from the ordinary Hamiltonian $H$ to the 
light-cone Hamiltonian $H-P^3$. 
In conventional light-cone calculations the theory is quantized directly 
on a light-like subspace (for reviews and further refs. see~\cite{lc}).
However, we always stay away from the light cone by having $r\neq 0$.
For studies concerning the possibility to approach the light cone 
continuously in the Hamiltonian method, we refer to~\cite{lcl}.
In Sect.~\ref{fi} below we will use the Lagrangian method and for us
the light-like limit concerns only the endpoint $r=0$ of the $r$ 
integration in Eq.~(\ref{tpme}).

Whereas for $r\in(0,2)$, the effective Hamiltonian $H_{eff}(r)$ has an 
energy gap, this gap vanishes for $r=0$ and 2. We have 
\be
\mbox{min}\left\{\langle X|H_{eff}(r)|X\rangle-E_0\right\}=\sqrt{r(2-r)}\,
M\,,\label{min}
\ee
where $E_0$ is the vacuum energy: 
\be
E_0=\langle 0|H_{eff}(r)|0\rangle=\langle 0|H|0\rangle\,,
\ee
and the minimum is taken over all normalized states $|X\rangle$ orthogonal 
to $|0\rangle$. Equation~(\ref{min}) is already indicative of a large 
correlation length and a critical phenomenon for $r=0$ and we will give 
more arguments for this in Sect.~\ref{ff}. 

Instead of representing $T_+(i\eta,Q^2,\mu)$ as an integral over a 
Euclidean matrix element ${\cal M}_E(y^0,r)$ in Eq.~(\ref{tpme}), 
we can also represent it as an integral over the corresponding
Minkowskian matrix element ${\cal M}_E(y^0,r)={\cal M}_M(-iy^0,r)$:
\be
T_+(i\eta,Q^2,\mu)=\frac{2i}{\sqrt{\eta^2-Q^2}}\int\limits^\infty_0dx^0
(x^0)^2\int\limits^2_0dr(1-r)\,e^{-x^0\left[-i\mu+(1-\frac{r}{2})\eta_-+
\frac{1}{2}r\eta_+\right]}\,{\cal M}_M(x^0,r).\label{tpmin}
\ee
with
\be
{\cal M}_M(x^0,r)=\langle P(p)|\,e^{+ix^0 H_{eff}(r)}\,J(0)\,e^{-ix^0 
H_{eff}(r)}\,J(0)|P(p)\rangle\,.\label{mme}
\ee
In Eq.~(\ref{tpmin}) there is no problem when setting $\mu=0$.

\section{Functional integral representation}\label{fi}
\setcounter{equation}{0}
In this section we will rewrite the matrix element of Eq.~(\ref{eme}) in 
terms of a path integral, using standard procedures as in the conventional 
case $H_{eff}=H$, i.e., $r=1$. We assume for simplicity that the states 
$|P(p)\rangle$ have the quantum numbers of the $\Phi$ field, i.e., that the 
$\Phi$ field is an interpolating field for these states. Then, according 
to the basic principles of the LSZ reduction formalism, we have to 
construct the operators 
\be
A(p,x^0)=i\int_{x^0=\mbox{\scriptsize const.}}d^3\vec{x}\,e^{ipx}\,\frac{
\stackrel{\leftrightarrow}{\partial}}{\partial x^0}\,\Phi(x)\,.
\ee
We have then, in the weak sense, 
\be
\lim_{x^0\to\pm\infty}\frac{1}{\sqrt{Z}}A^\dagger(p,x^0)|0\rangle=|P(p)
^{out}_{in}\rangle=|P(p)\rangle\,,\label{lsz}
\ee
where $Z$ is the wave function renormalization constant. 

The Euclidean version of Eq.~(\ref{lsz}) for the proton with zero 
three-momentum reads:
\be
\frac{1}{\sqrt{Z}}\lim_{t\to\infty}\left(e^{-Ht}A^\dagger(0)|0\rangle
\right)e^{(M+E_0)t}=|P(p)\rangle\,,\label{lsze}
\ee
where we set $A(0)\equiv A(p,0)$ with $p=(M,\vec{0})$. In terms 
of the field operator and its conjugate canonical momentum, we have 
\bea
A(0)&=&\int_{x^0=0}d^3\vec{x}\left(i\Pi(x)+M\Phi(x)\right)\,,\label{adef}\\
A^\dagger(0)&=&\int_{x^0=0}d^3\vec{x}\left(-i\Pi(x)+M\Phi(x)\right)\,.
\label{addef}
\eea

Before rewriting the matrix element of Eq.~(\ref{eme}) in terms of a path 
integral, let us recall the corresponding procedure in the conventional 
case $H_{eff}=H$, i.e., $r=1$. Using Eq.~(\ref{lsze}) we have 
\be
{\cal M}_E(\tau,1)\!=\!\frac{1}{Z}\lim_{\tau_i\to-\infty\atop\tau_f\to+
\infty}\,e^{(\tau_f-\tau_i)(M+E_0)}\,\langle 0|A(0)\,e^{-(\tau_f-\tau)H}\,
\!J(0)\,e^{-(\tau-0)H}\,\!J(0)\,e^{-(0-\tau_i)H}\,A^\dagger(0)|0\rangle\,.
\label{emef}
\ee
Now the standard procedure is to introduce alternating intermediate states 
which are eigenstates of the field operators $\Pi$ and $\Phi$ with definite 
classical field eigenvalues $\pi$ and $\phi$ on a sufficiently dense grid 
in Euclidean time. The $\pi$ integrations can then be carried out 
explicitly, leaving one with a product of $\phi$ integrations. In the limit 
where the time slices become arbitrarily thin, this defines the Euclidean 
path integral. 

In realizing this procedure, all operators in Eq.~(\ref{emef}) are replaced 
by functionals of $\pi$ and $\phi$. For $H$ and $J$, the relevant 
expressions are given in Eqs.~(\ref{hop}) and (\ref{je}), for the 
annihilation and creation operator $A$, $A^\dagger$ in Eqs.~(\ref{adef}) 
and (\ref{addef}). 

The standard procedure outlined above gives the result 
\be
{\cal M}_E(\tau,1)=\frac{1}{Z}\lim_{\tau_i\to-\infty\atop\tau_f\to+\infty}
\,e^{(\tau_f-\tau_i)M}\,{\cal Z}^{-1}\int D\phi\, a(\tau_f)j(\tau)j(0)
a^\dagger(\tau_i)\,e^{-\int d^4x\,{\cal L}_E}\,,\label{emep}
\ee
\be
{\cal Z}=\int D\phi\,e^{-\int d^4x\,{\cal L}_E}\,,
\ee
with the Euclidean Lagrangian
\bea
{\cal L}_E&=&-i\pi(x)\dot{\phi}(x)+{\cal H}\\
&=&-i\pi(x)\dot{\phi}(x)+\frac{1}{2}\pi(x)^2+\frac{1}{2}\Big(\nabla\phi(x)
\Big)^2+\frac{m^2}{2}\phi(x)^2+\frac{\lambda}{4!}\phi(x)^4\,.\nonumber
\eea
Here the canonical momentum $\pi$ entering Eq.~(\ref{emep}) via $a$, 
$a^\dagger$ and ${\cal L}_E$ has to be understood as a function of $\phi$. 
In the Euclidean theory, the relation between $\phi$ and $\pi$ is defined 
by 
\be
i\dot{\phi}=\frac{\partial{\cal H}}{\partial\pi}=\pi\,.\label{pf}
\ee
In this way, Eq.~(\ref{emep}) and the expressions for $a$ and ${\cal L}_E$ 
follow directly from Eq.~(\ref{emef}) and the explicit form of the 
Hamiltonian density ${\cal H}$. 

The result at $r\neq 1$ can be obtained by following the standard procedure 
for converting a matrix element into a functional integral expression 
outlined above. It is easy to guess the correct answer by observing that 
the effective Hamiltonian density is defined by the substitution 
\be
{\cal H}\,\,\longrightarrow\,\,{\cal H}_{eff}\,=\,{\cal H}+(1-r)\pi
\partial_3\phi\,.
\ee
Thus, the expression for $\pi$ in terms of $\phi$ following from 
Eq.~(\ref{pf}) now reads 
\be 
i\dot{\phi}=\frac{\partial{\cal H}_{eff}}{\partial\pi}=\pi+(1-r)\,
\partial_3\phi\,.
\ee
This is to be used when expressing $a$, $a^\dagger$ and the new effective 
Lagrangian ${\cal L}_{E,\,eff}$ in terms of $\phi$ in Eq.~(\ref{emep}). 

We summarize the results of this calculation, some details of which are given 
in Appendix~\ref{fic}: the Euclidean matrix element of Eq.~(\ref{eme}) can 
be calculated as 
\be
{\cal M}_E(\tau,r)=\frac{1}{Z}\lim_{\tau_i\to-\infty\atop\tau_f\to+\infty}
\,e^{(\tau_f-\tau_i)M}\,{\cal Z}^{-1}\int D\phi\, a(\tau_f)j(\tau)j(0)
a^\dagger(\tau_i)\,e^{-\int d^4x\,{\cal L}_{E,\,eff}}\,,\label{emer}
\ee
\be
{\cal Z}=\int D\phi\,e^{-\int d^4x\,{\cal L}_{E,\,eff}}\,,
\ee
with
\be
{\cal L}_{E,\,eff}=\frac{1}{2}\Big(\partial_0{\phi}(x)+i(1-r)\,\partial_3
\phi(x)\Big)^2+\frac{1}{2}\Big(\nabla\phi(x)\Big)^2+\frac{m^2}{2}\phi(x)^2+
\frac{\lambda}{4!}\phi(x)^4\,\label{el}
\ee
and
\bea
a(\tau_f)&=&\int_{x^0=\tau_f}d^3\vec{x}\,\Big(-\partial_0{\phi}
(x)-i(1-r)\,\partial_3\phi(x)+M\phi(x)\Big)\,,\label{af}\\
a^\dagger(\tau_i)&=&\int_{x^0=\tau_i}d^3\vec{x}\,\Big(\partial_0{\phi}
(x)+i(1-r)\,\partial_3\phi(x)+M\phi(x)\Big)\,,\label{afd}
\eea
\be
j(\tau)=\phi^2(\vec{0},\tau)\,.
\ee
Note that in the Euclidean path integral, Eq.~(\ref{emer}), the functionals 
$a(\tau)$ and $a^\dagger(\tau)$ are not complex conjugate to each other and 
${\cal L}_{E,\,eff}$ is not a real function for $r\neq 1$. 

The matrix element Eq.~(\ref{emer}) is related to the high-energy limit of 
the structure function $W$ via Eqs.~(\ref{imt}), (\ref{trtp}) and 
(\ref{tpme}). The above Euclidean functional integral expression for 
${\cal M}_E$ is one of the central results of this paper. We propose to 
evaluate it using genuinely non-perturbative methods, e.g., lattice Monte 
Carlo simulations. 

Similarly, the Minkowskian matrix element of Eq.~(\ref{mme}) can be 
expressed in terms of an appropriate Minkowskian functional integral 
(cf.~Appendix~\ref{fic}).

\section{Free-field example}\label{ff}
\setcounter{equation}{0}
To illustrate the meaning of the formalism developed, let us consider the 
simple example of free fields, i.e., the case $\lambda=0$. 

\begin{figure}[ht]
\begin{center}
\vspace*{.2cm}
\parbox[b]{6.5cm}{\psfig{width=6.5cm,file=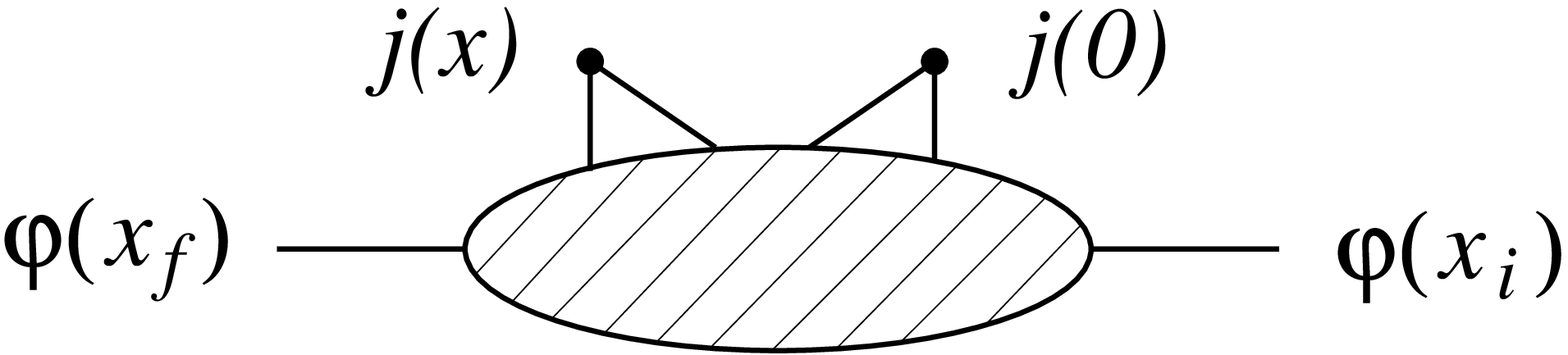}}\\
\end{center}
\refstepcounter{figure}\label{gf}
{\bf Figure \ref{gf}:} The Euclidean Green function required for the 
calculation of ${\cal M}_E$. 
\end{figure}

Using Eqs.~(\ref{emef}), (\ref{af}) and (\ref{afd}), the matrix element is 
expressed as 
\be
{\cal M}_E(\tau,r)=\frac{1}{Z}\lim_{\tau_i\to-\infty\atop\tau_f\to+\infty}
\,e^{(\tau_f-\tau_i)M}\,\int d^3\vec{x}_f\,d^3\vec{x}_i\left(
-\frac{\partial}{\partial\tau_f}+M\right)\left(\frac{\partial}{\partial
\tau_i}+M\right)G_{\phi jj\phi}(x_f,x,0,x_i)\,,\label{emeg}
\ee
where 
\be
x_i=(\vec{x}_i,\tau_i)\quad,\quad x_f=(\vec{x}_f,\tau_f)\quad,\quad 
x=(\vec{0},\tau)\,,
\ee
and $G_{\phi jj\phi}$ is the connected Green function of two fields $\phi$ 
and two currents $j$ illustrated in Fig.~\ref{gf}. The terms with $x^3$ 
derivative disappear because of the $x^3$ integration. 

\begin{figure}[ht]
\begin{center}
\vspace*{.2cm}
\parbox[b]{9cm}{\psfig{width=9cm,file=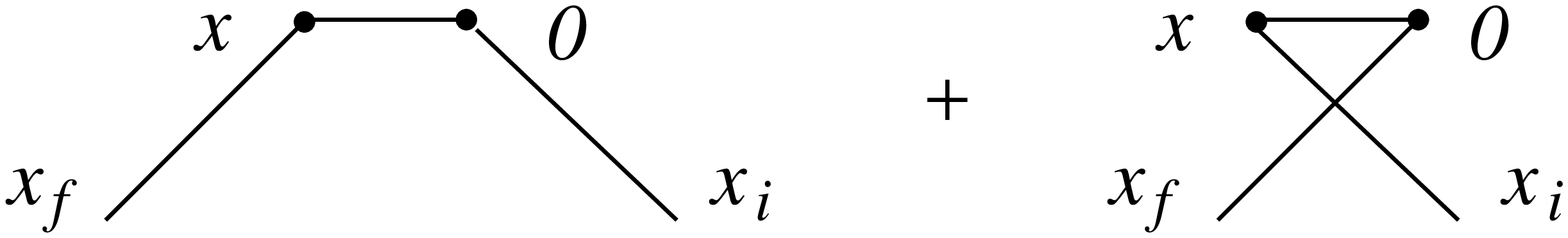}}\\
\end{center}
\refstepcounter{figure}\label{td}
{\bf Figure \ref{td}:} The two diagrams contributing to the Euclidean 
Green function $G_{\phi jj\phi}$ in the free-field case. 
\end{figure}

In the free-field case, Fig.~\ref{gf} contains only the two diagrams shown 
in Fig.~\ref{td}, where the propagator following from the free part of the
Euclidean Lagrangian, Eq.~(\ref{el}), has to be used. It reads
\be
G(x)=\int\frac{d^4k}{(2\pi)^4}\,e^{ikx}\,\frac{1}{k^TA_rk+m^2}=\frac{m}
{4\pi^2\sqrt{x^TA_r^{-1}x}}\,K_1\left(m\sqrt{x^TA_r^{-1}x}\right)\,,
\label{egf}
\ee
where 
\be
A_r=\left(\begin{array}{cccc}1&0&0&0\\0&1&0&0\\0&0&1-(1-r)^2&i(1-r)\\
0&0&i(1-r)&1\end{array}\right)\quad\mbox{with}\quad
x=\left(\begin{array}{c}x^1\\x^2\\x^3\\\tau\end{array}\right)\quad,\quad
k=\left(\begin{array}{c}k_1\\k_2\\k_3\\k_4\end{array}\right)\,.
\ee
$K_1$ is the modified Bessel function, and $m=M$ in our free-field model. It 
is easy to see that $x^TA_r^{-1}x\simeq 2r\tau^2$ for $x=(\vec{0},\tau)$ 
and $r\to 0$. According to Eq.~(\ref{egf}), this means that the correlation 
length in the Euclidean time direction diverges and we are dealing with a 
critical phenomenon. 

To study in general the behaviour of $G(x)$ for $x\to\infty$ we set 
\be
x=R\hat{x}\qquad,\qquad R=|x|=\sqrt{x_1^2+x_2^2+x_3^2+\tau^2}
\ee
and consider the limit $R\to\infty$ for fixed unit vector $\hat{x}$. We get 
from Eq.~(\ref{egf}):
\be
G(x)\longrightarrow\frac{1}{4\pi^2}\sqrt{\frac{\pi}{2}}\,M^2\left(M^2R^2
\hat{x}^TA_r^{-1}\hat{x}\right)^{-3/4}\!e^{-MR\sqrt{\hat{x}^TA_r^{-1}
\hat{x}}}\qquad\mbox{for}\qquad{R\to\infty}\,.
\ee
Thus the decay of $|G(x)|$ as $R\to\infty$ is governed by exp$\left[-MR\,
\mbox{Re}\sqrt{\hat{x}^TA_r^{-1}\hat{x}}\,\right]$. This suggests to define 
a correlation length depending on $\hat{x}$ as
\be
\xi(\hat{x})=\left[M\,\mbox{Re}\sqrt{\hat{x}^TA_r^{-1}\hat{x}}\right]^{-1}
\,.
\ee
We have, in the Euclidean time direction, 
\be
\xi(\vec{0},1)=\left[M\sqrt{r(2-r)}\right]^{-1}\label{xir} ~,
\ee
and, in the space directions,
\be
\xi(\vec{e}_j,0)=M^{-1}\,.
\ee
Thus for $r\to 0$ the correlation in this free-field model becomes of long 
range only in the time direction. There $\xi(\vec{0},1)$ diverges as 
$1/\sqrt{r}$, i.e., in a manner characteristic of mean-field theories for 
critical phenomena in statistical mechanics, identifying $r$ with $T-T_c$, 
the deviation of the temperature from the critical value. 

We also note the behaviour of $G(x)$, Eq.~(\ref{egf}), for $R\to 0$: 
\be
G(x)\longrightarrow\frac{1}{4\pi^2}\left(R^2\hat{x}A_r^{-1}\hat{x}
\right)^{-1}\qquad\mbox{for}\qquad R\to 0\,.
\ee
In the Euclidean time direction this can be written as 
\be
G(\vec{0},\tau)\longrightarrow\frac{M^2}{4\pi^2}\left(\frac{\xi(\vec{0},1)}
{\tau}\right)^2\qquad\mbox{for}\qquad \tau\ll\xi(\vec{0},1)\,.\label{smt}
\ee
This means that for times much smaller than the correlation length the 
Green function in the free theory has a simple power behaviour, again 
similar to what is found in the study of critical phenomena. 

\begin{figure}[ht]
\begin{center}
\vspace*{.2cm}
\parbox[b]{11cm}{\psfig{width=11cm,file=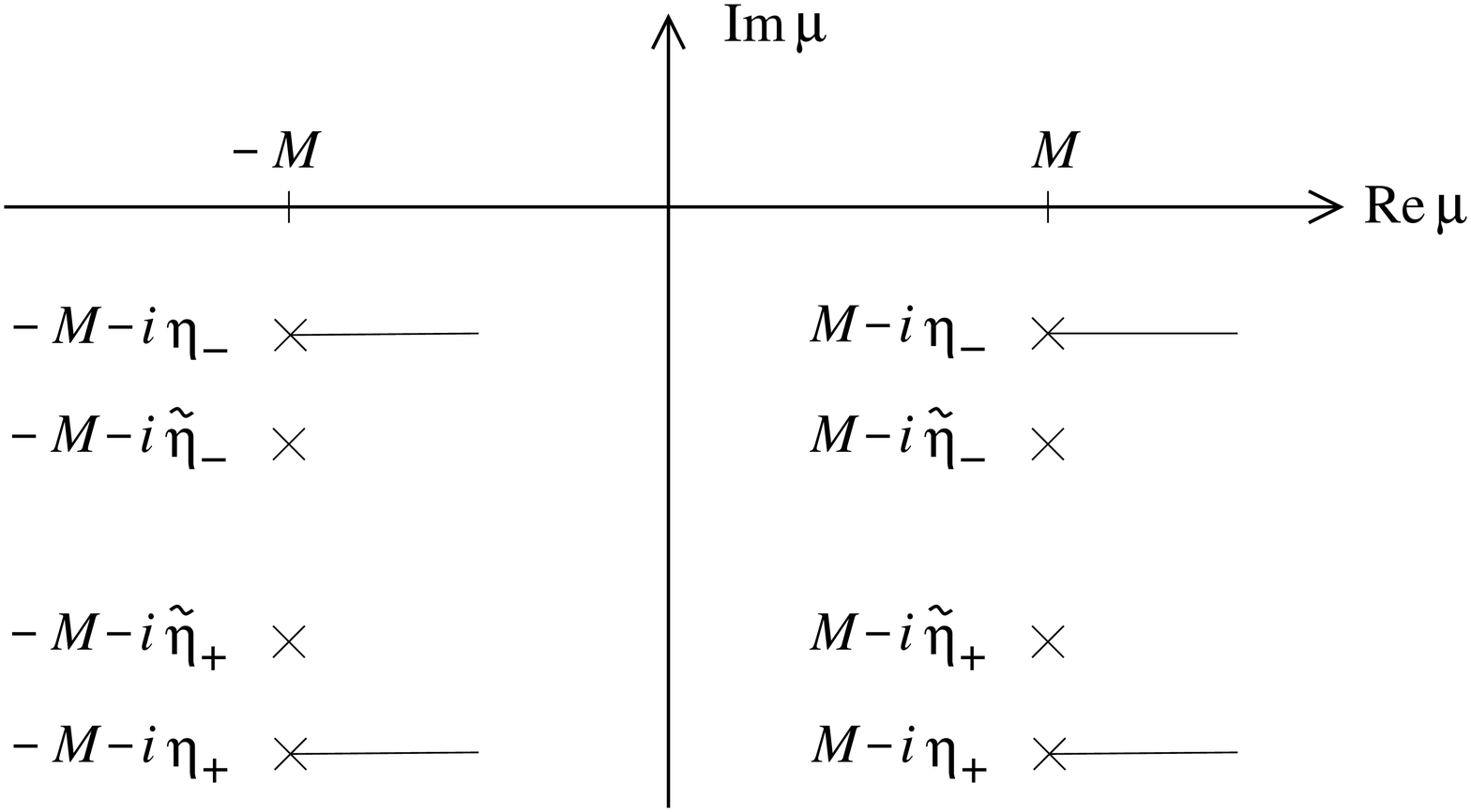}}\\
\end{center}
\refstepcounter{figure}\label{sing}
{\bf Figure \ref{sing}:} The singularities of $T_+(i\eta,Q^2,\mu)$ in the 
complex $\mu$ plane. There are poles at $\pm M-i\tilde{\eta}_+$, $\pm M 
-i\tilde{\eta}_-$ with $\tilde{\eta}_\pm=\eta\pm\sqrt{\eta^2-Q^2-M^2}$ and 
logarithmic branch cuts starting at $\pm M-i\eta_+$ and $\pm M-i\eta_-$. 
Here we assume $\eta^2>Q^2+M^2$. 
\end{figure}

Now we describe the calculation of $T_+$ in Eq.~(\ref{tpme}) in the free-field 
case. Changing the integration variables in Eq.~(\ref{tpme}) from $y_\pm$ 
to $y^0$ and $y^3$, using the expression of Eq.~(\ref{emeg}) for ${\cal 
M}_E$, and calculating the Green function $G_{\phi jj\phi}$ according to 
Fig.~\ref{td}, the amplitude $T_+$ reads 
\bea
T_+(i\eta,Q^2,\mu)&=&\frac{-1}{\pi^2\sqrt{\eta^2-Q^2}}\int_0^\infty dy^0
\int_{-y^3}^{y^3}dy^3\,y^3\,e^{\frac{1}{2}[(y^0+y^3)(\mu+i\eta_-)+
(y^0-y^3)(\mu+i\eta_+)]}\nonumber
\\
&&\times\left[e^{My^0}+e^{-My^0}\right]\frac{M}{\sqrt{(y^0)^2-(y^3)^2}}
K_1\left(M\sqrt{(y^0)^2-(y^3)^2}\right)\,.
\eea
Applying the identity
\be
\frac{y^3M}{\sqrt{(y^0)^2-(y^3)^2}}K_1\left(M\sqrt{(y^0)^2-(y^3)^2}\right)
=\frac{\partial}{\partial y^3}K_0\left(M\sqrt{(y^0)^2-(y^3)^2}\right)\,,
\ee
integrating by parts, and returning to the variables $y_\pm$, one finds 
\bea
T_+(i\eta,Q^2,\mu)&=&\frac{-i}{4\pi^2}\int_0^\infty dy_+
\int_0^\infty dy_-\,e^{\frac{1}{2}[y_+(\mu+i\eta_-)+y_-(\mu+i\eta_+)]}
\nonumber
\\
&&\times\left[e^{M(y_++y_-)/2}+e^{-M(y_++y_-)/2}\right]K_0\left(M
\sqrt{y_+y_-}\right)\,.
\eea
For sufficiently negative Re$\,\nu$, the integrals can be evaluated using 
the relations 6.631.3, 9.224 and 6.224.1 of~\cite{gr}. The result is 
\be
T_+(i\eta,Q^2,\mu)=-\frac{i}{2\pi^2}\,\Big\{f(M)+f(-M)\Big\}
\ee
with
\be 
f(M)=\frac{1}{\mu^2+2M\mu-Q^2+2i\eta(M\!+\!\mu)}\ln\left[\frac{1}{M^2}
\left\{(M\!+\!\mu)^2-Q^2+2i\eta(M\!+\!\mu)\right\}\right].
\ee
When this expression is analytically continued to $\mu=0$, a careful 
treatment of the imaginary part of the logarithms is essential. The 
singularities of $T_+(i\eta,Q^2,\mu)$ in the complex $\mu$ plane are shown 
in Fig.~\ref{sing}. The continuation in $\mu$ from Re$\,\mu<-M$ to $\mu=0$ 
has to be done along the real axis staying above the left hand 
singularities. 

{}From Eq.~(\ref{trtp}) one finds the behaviour of the retarded amplitude 
on the imaginary axis in the $\nu$ plane: 
\be
T_r(i\eta,Q^2)=\mbox{Re}\,T_+(i\eta,Q^2,0)=\frac{1}{2\pi}\left(\frac{1}
{Q^2-2i\eta M}+\frac{1}{Q^2+2i\eta M}\right)\,.
\ee
The corresponding behaviour on the real axis follows from the analytic 
continuation $i\eta\to\nu$ for $\nu$ real:
\be
T_r(\nu,Q^2)=\frac{1}{2\pi}\left(\frac{1}{Q^2-2\nu M-i\varepsilon}+\frac{1}
{Q^2+2\nu M+i\varepsilon}\right)\,.
\ee
This is in agreement with what one would have found by simply adding the 
two diagrams for the forward scattering of a scalar proton and a scalar photon 
in the original Minkowskian theory. 

We will close this section with some speculative remarks. Looking at 
Eq.~(\ref{tpme}) we see that for $\eta\to\infty$ the typical integration 
range over ${\cal M}_E(y^0,r)$ is 
\be
0\,\,{\scriptstyle\leq}\,\,r\,\mst{<}{\sim}\,\frac{1}{\eta y^0}\,\,,
\label{rv}
\ee
\be
0\,\,{\scriptstyle\leq}\,\,y^0\,\mst{<}{\sim}\,\frac{2\eta}{Q^2}\equiv 
y^0_m\,.
\ee
Let us insert the mean value of $y^0$ of Eq.~(\ref{tpme}) into 
Eq.~(\ref{rv}) to get as typical value for $r$:
\be
\bar{r}=\frac{Q^2}{\eta^2}\,.
\ee
Then we have two relevant scales for the $y^0$ integration in 
Eq.~(\ref{tpme}): $y^0_m$ and the length scale of the decay of ${\cal M}_E 
(y^0,\bar{r})$. Also in the general case of the theory with interactions 
this should be governed by a correlation length $\xi(\bar{r})$ similar to 
$\xi^{(0)}(\bar{r})\equiv\xi(\vec{0},1)$, Eq.~(\ref{xir}), in the free-field 
case. Clearly the behaviour of the amplitude $T_+$ will depend 
crucially on whether $y^0_m$ is smaller or bigger than $\xi(\bar{r})$. 

In the free-field case we have 
\be
y^0_m\,\,\mst{<}{>}\,\,\xi^{(0)}(\bar{r})\qquad\mbox{for}\qquad Q^2\,\,
\mst{>}{<}\,\,8M^2\,.
\ee
Here for large $Q^2$ and $\eta\to\infty$ the integration in 
Eq.~(\ref{tpme}) probes only the region of ${\cal M}_E(y^0,r)$ where $y^0$ 
is smaller than the correlation length and thus where Eq.~(\ref{smt}) 
applies. We can speculate that in the theory with interaction we should 
again be able to distinguish two regimes,
\be
y^0_m\,\,\mst{<}{>}\,\,\xi(\bar{r})\,,\label{tre}
\ee
where, from our experience with critical phenomena, we would expect 
$\xi(\bar{r})$ to satisfy a scaling law for $\bar{r}\to 0$,
\be
\xi(\bar{r})\simeq\frac{c}{M}(\bar{r})^{-\rho}\,,\qquad c=\mbox{const.}\,,
\qquad\rho>0\,,
\ee
but not with the mean field exponent $\rho=1/2$ as in Eq.~(\ref{xir}). 

For $0<\rho<1$ the dividing line between the two regimes of Eq.~(\ref{tre}) 
would then be roughly:
\be
\eta=\frac{2M}{c}\left(\frac{Qc}{2M}\right)^{\frac{\scriptstyle 2-2\rho}
{\scriptstyle 1-2\rho}}\label{dl}
\ee
and it is tempting to speculate that this could correspond to the dividing 
line between hard $[y^0_m<\xi(\bar{r})]$ and soft $[y^0_m>\xi(\bar{r})]$ 
pomeron regimes. But much more work is needed before such a conjecture can 
be substantiated. 

\begin{figure}[ht]
\begin{center}
\vspace*{.2cm}
\parbox[b]{11.5cm}{\psfig{width=11.5cm,file=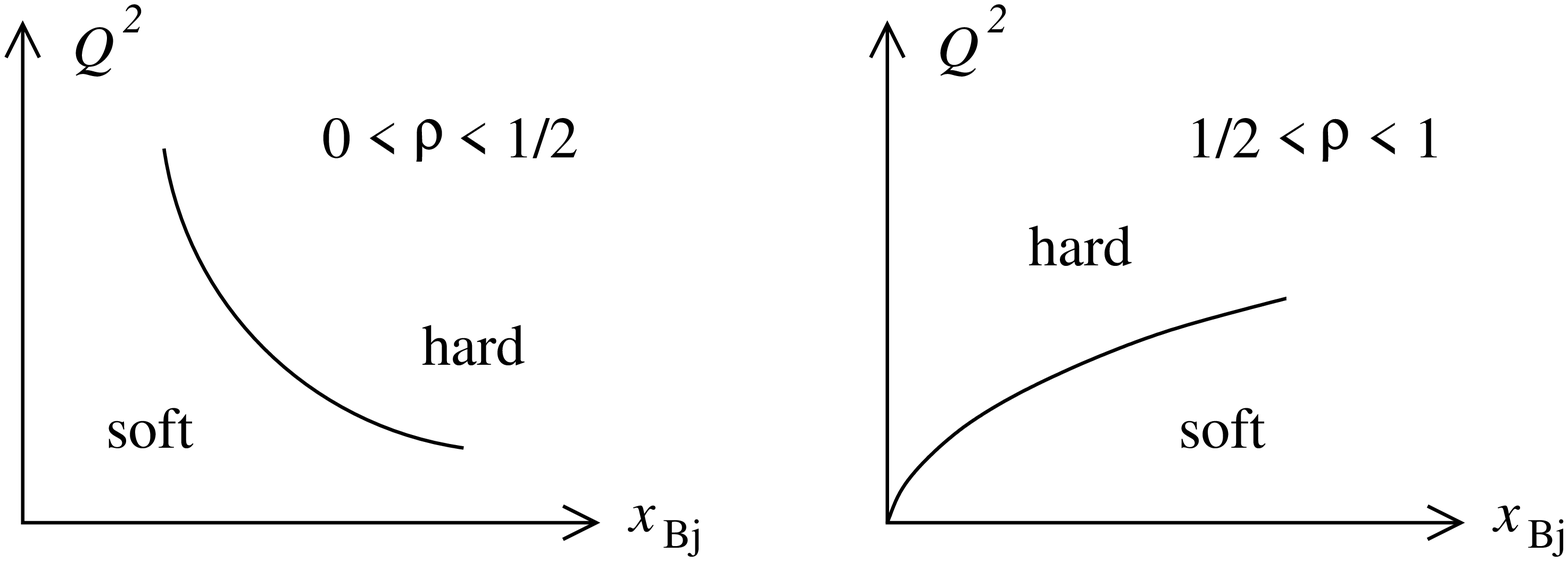}}\\
\end{center}
\refstepcounter{figure}\label{q2x}
{\bf Figure \ref{q2x}:} Sketch of the regions of soft and hard pomeron 
effects in the $\xbj$--$Q^2$ plane using Eq.~(\ref{q2xf}).
\end{figure}

Let us, nevertheless, assume for the moment that such a conjecture is true 
and apply Eq.~(\ref{dl}) also in the physical region, replacing $\eta$ by 
$\nu$. Then  we should find a dividing line between hard and soft pomeron 
effects in the structure functions in the $\xbj$ versus $Q^2$ plane: 
\be
Q^2=\frac{4M^2}{c^2}(c\xbj)^{2-\frac{\scriptstyle 1}{\scriptstyle \rho}}\,.
\label{q2xf}
\ee
This is illustrated in Fig.~\ref{q2x}. Thus, taking the limit $\xbj\to 0$ 
at fixed $Q^2>0$ we would finally always go from the hard to the soft 
regime for $0<\rho<1/2$ and vice versa for $1/2<\rho<1$. 

\section{Alternative analytic continuations to Euclidean space}\label{alt} 
\setcounter{equation}{0}
It is clear that the form of the analytic continuation from the Minkowski 
to the Euclidean theory discussed in the previous sections is by no means 
unique. Other possible forms of analytic continuations from Minkowski to 
Euclidean space can be derived: in this section we discuss two of these 
alternative Euclidean formulations and compare them with the one of 
Sect.~\ref{mieu}. 

Working in the rest frame of the proton, consider the quantity 
\be
T(q^0,\vec{q}\,) = {i \over 2\pi} \int d^4x\,\theta (x^0) e^{iqx}
\langle P(p) \vert J(x) J(0) \vert P(p) \rangle\label{tdef}
\ee
as an analytic function of $q^0$ with the spatial part $\vec{q}$ fixed and 
real. Using standard methods, one finds
\be
T(q^0,\vec{q}\,) = -\sum_X(2\pi)^2 \delta^{(3)} (\vec{q} - \vec{p}\,')
{1 \over q^0 +M - p'^0}\vert \langle P(p) \vert J(0) \vert X(p') \rangle 
\vert^2\,,\label{nt}
\ee
which reveals the cut along the positive real axis. The structure function 
of Eq.~(\ref{sf}) is related to $T$ by 
\be
{\rm Im} [ T(q^0 + i\varepsilon,\vec{q}\,) ] = W(\nu,Q^2) ~,
\ee
where $\varepsilon \to 0+$.
Now, consider $T(q^0,\vec{q}\,)$ on the positive imaginary axis, i.e., for 
$q^0 = i\eta$ with $\eta > 0$. Using the identity 
\be
{1 \over i\eta+M-p'^0}=-\int_0^{+\infty} dx_4
e^{(i\eta +M-p'^0)x_4}
\ee
(where the problematic lowest value of $p'^0$, $p'^0=M$, can be treated by 
giving $\eta$ a small positive imaginary part), the following expression 
for $T$ is obtained from Eq.~(\ref{nt}):
\be
T(i\eta,\vec{q}\,) = {1 \over 2\pi} \displaystyle\int d^3\vec{x}
\int_0^{+\infty} dx_4 e^{i(\eta x_4 - \vec{q} \cdot \vec{x})}\langle P(p) 
\vert J_E(\vec{x},x_4) J_E(0) \vert P(p) \rangle ~.\label{tje}
\ee
Here $J_E(\vec{x},x_4)$ is the current of the conventional Euclidean 
theory: 
\be
J_E(\vec{x},x_4) = e^{H x_4} J(0,\vec{x}) e^{-H x_4} ~.
\label{jedef}
\ee
Using the methods of Sect.~\ref{camp} [cf.~Eqs.~(\ref{rot}) and 
(\ref{trm})], invariance under spatial rotations can be employed to 
further simplify the above expression for $T$. 

To summarize, the amplitude $T$ on the real axis, which determines the 
physical structure function, can be obtained from the one on the imaginary 
axis by analytic continuation in $q^0$: $(q^0=i\eta)\,\to\,(q^0=\nu)$. 
According to Eq.~(\ref{tje}), the latter one can be calculated from a 
conventional Euclidean matrix element. 

This approach is simpler than the one discussed in the previous sections 
in that the Euclidean theory is based on the usual Hamiltonian $H$ and not 
on the more complicated effective Hamiltonian of Eq.~(\ref{heff}). However, 
in order to derive the large-$\nu$ behaviour of the structure function 
$W(\nu,Q^2)$ at fixed $Q^2$, one needs the analytic continuation of 
$T(q^0,\vec{q}\,)$ in $q^0$ for every value of $\vec{q}$. Only at this 
point can one study the asymptotic limit $\nu \to \infty$, which involves 
both $q^0\to\infty$ and $|\vec{q}\,|\to\infty$ in the physical region. This 
analytic continuation in $q^0$ is expected to be more troublesome than 
the $\mu$ continuation of Sect.~\ref{camp}, which is decoupled from the 
high-energy limit. 

We consider now a third possible form of analytic continuation from 
Minkowski to Euclidean space, which is close in spirit to~\cite{meg}. 
The basic object is still the virtual Compton amplitude Eq.~(\ref{tr}), 
which we write now as: 
\be
T_r(\nu,Q^2)=\frac{1}{2}\left(\tilde{T}_+(q)+\tilde{T}_+^*(-q^*)\right)\, 
\label{ntr}
\ee
where
\be
\tilde{T}_+(q)=\frac{i}{\pi}\int d^4x\,\theta(x^0)\theta((x^0)^2-(x^3)^2)
\,e^{iqx}\,\langle P(p)|J(x)J(0)|P(p)\rangle\,.
\ee
Here we work in the rest system of the proton and choose now $q$ in the 
form
\be
q = (\nu,\vec{e}_\perp\sqrt{\nu^2 + Q^2}, 0)\,,
\ee
with $\vec{e}_\perp$ a unit vector in the $(q^1,q^2)$ plane. With the 
parametrization $x=(\xi\cosh\chi,\vec{x}_\perp,\xi\sinh\chi)$, we find 
\be
\tilde{T}_+(q)={i\over\pi}\int d^2\vec{x}_\perp\,e^{-i\vec{e}_\perp \cdot
\vec{x}_\perp\sqrt{\nu^2+Q^2}}\int_{-\infty}^{+\infty}d\chi\int_0^{+\infty}
d\xi \xi\,e^{i\nu\xi\cosh\chi}\langle P(p)\vert J(x)J(0)\vert P(p)
\rangle ~.
\ee
We will show how to compute this quantity and its $\nu$ derivatives at 
$\nu = 0$ using an analytic continuation from Minkowski to Euclidean 
space. First of all, we write
\be
\tilde{T}_+(q)\vert_{\nu =0}=\int d^2\vec{x}_\perp\,e^{-i\vec{q}_\perp \cdot 
\vec{x}_\perp}\int_{-\infty}^{+\infty} d\chi A^{(1)}(\chi,\vec{x}_\perp) ~,
\ee
where $\vec{q}_\perp=Q\vec{e}_\perp$ and the function $A^{(1)}$ is defined 
by 
\ba
A^{(1)}(\chi,\vec{x}_\perp)&=&{i\over\pi}\int_0^{+\infty}d\xi\xi\langle 
P(p)\vert J(x) J(0) \vert P(p) \rangle \nonumber 
\\
&=& {-i \over\pi} \sum_X\langle P(p) \vert J(0,\vec{x}_\perp,0) \vert 
X(p') \rangle \langle X(p') \vert J(0) \vert P(p) \rangle \nonumber 
\\ 
& & \times {1 \over \left[ (p'^0 - M) \cosh\chi - p'^3 \sinh\chi - i
\varepsilon\right]^2} ~,
\ea
where finally the limit $\varepsilon \to 0+$ is taken. The singularity 
structure in the complex $\chi$ plane is such that the function can be 
continued to the imaginary axis, $\chi\to i\theta$ where, however, the 
$\theta$ range is restricted by $\theta\in(-\pi/2,\pi/2)$. In this region, 
$A^{(1)}$ is related to a Euclidean matrix element,
\ba
A^{(1)}(i\theta,\vec{x}_\perp) &=&
{-i \over\pi} \displaystyle\int_0^{\infty} d\xi \xi 
\langle P(p) \vert J_E(x_E) J_E(0) \vert P(p) \rangle\nonumber \\
&=& {-i \over\pi} \displaystyle\sum_X
\langle P(p) \vert J(0,\vec{x}_\perp,0) \vert X(p') \rangle 
\langle X(p') \vert J(0) \vert P(p) \rangle \nonumber \\ 
& & \times {1 \over \left[ (p'^0 - M) \cos\theta - ip'^3 \sin\theta
-i\varepsilon\right]^2} ~,
\ea
where $x_E=(\vec{x}_\perp, \xi \sin\theta, \xi \cos\theta)$ is a Euclidean 
four-vector and the Euclidean current $J_E(x_E)$ is defined by 
Eq.~(\ref{jedef}). 

More generally, the expression for the $n$-th derivative of $\tilde{T}_+$ 
with respect to $\nu$, evaluated at $\nu = 0$, reads
\be
\partial^n_\nu\tilde{T}_+(q)\vert_{\nu =0}=\sum_{l=0}^n\int d^2
\vec{x}_\perp e^{-i\vec{q}_\perp \cdot \vec{x}_\perp} C_l^n(\vec{q}_\perp\!
\cdot\!\vec{x}_\perp,Q^2)\int_{-\infty}^{+\infty} d\chi (\cosh\chi)^l A^{(l+1)}
(\chi,\vec{x}_\perp) ~,\label{cl}
\ee
with explicitly calculable coefficients $C_l^n$ and functions $A^{(l)}$ 
defined by
\ba
A^{(l)}(\chi,\vec{x}_\perp)&=&{i\over\pi}\int_0^{\infty} d\xi \xi^l
\langle P(p) \vert J(x) J(0) \vert P(p) \rangle\nonumber \\
& & = {(-i)^l l! \over\pi}\sum_X\langle P(p) \vert J(0,\vec{x}_\perp,0) 
\vert X(p') \rangle\langle X(p') \vert J(0) \vert P(p) \rangle \nonumber \\
&&\times {1 \over \left[ (p'^0 - M) \cosh\chi - p'^3 \sinh\chi - i\varepsilon
\right]^{l+1}} ~.
\ea
As in the case $l = 1$, the values of this function for real $\chi$ can be 
obtained from 
\ba
A^{(l)}(i\theta,\vec{x}_\perp)&=&{(-i)^l \over\pi} \int_0^\infty d\xi
\xi^l\langle P(p) \vert J_E(x_E) J_E(0) \vert P(p) \rangle\nonumber \\
& & = {(-i)^l l! \over\pi} \sum_X\langle P(p) \vert J(0,\vec{x}_\perp,0) 
\vert X(p') \rangle \langle X(p') \vert J(0) \vert P(p) \rangle \nonumber\\
&& \times {1 \over \left[ (p'^0 - M) \cos\theta - ip'^3 \sin\theta
-i\varepsilon \right]^{l+1}} ~.
\ea
by the analytic continuation $i\theta \to \chi$. 

To summarize, one first computes $A^{(l)}(i\theta,\vec{x}_\perp)$ for $l=1, 
\ldots, n+1$ and $\theta\in(-\pi/2,\pi/2)$ in the Euclidean field theory (for 
example, on the lattice). Then, the value of $\partial^n_\nu \tilde{T}_+(q)
\vert_{\nu = 0}$ can be obtained from Eq.~(\ref{cl}) using the analytic 
continuations of the $A^{(l)}$ to the real axis. Thus, using 
Eq.~(\ref{ntr}) and a Taylor expansion of $\tilde{T}_+$ around $\nu = 0$, 
the behaviour of $T_r(\nu,Q^2)$ at small $\nu$ (corresponding to large 
$\xbj$) can be studied. The values of $\partial^n_\nu T_r(\nu,Q^2)\vert_{
\nu = 0}$ can then be related to integrals of the structure function 
$W(\nu, Q^2)$ using dispersion relations in $\nu$. Assuming for simplicity 
no subtractions this reads:
\be
T_r(\nu,Q^2)=\frac{1}{\pi}\int_{Q^2/2M}^\infty d\nu'\,W(\nu',Q^2)\left[
\frac{1}{\nu'-\nu-i\varepsilon}+\frac{1}{\nu'+\nu+i\varepsilon}\right]\,,
\ee
\be
\partial^n_\nu T_r(\nu,Q^2) \vert_{\nu = 0}=\frac{n!}{\pi}
\int_{Q^2/2M}^\infty d\nu' {W(\nu', Q^2) \over (\nu')^{n+1}}[1+(-1)^n]\,,
\qquad (n=0,1,2,...)\,.
\ee
Again, the advantage with respect to the method of Sect.~\ref{mieu} is 
the use of the conventional Hamiltonian in the Euclidean theory. The 
disadvantage is the limited sensitivity to the region of small $\xbj$.

\section{Conclusions}\label{conc}
\setcounter{equation}{0}
In this paper, a fundamentally new approach to the long-standing problem of 
small-$\xbj$ structure functions in DIS has been developed. It is based on 
the well-known relation of the small-$\xbj$ limit of DIS and the high-energy 
limit of forward virtual Compton scattering. Instead of taking the energy 
of the forward virtual Compton amplitude to infinity along the real axis, 
we propose to consider the limit of large imaginary energy. According to 
the theorems of Phragm\'{e}n and Lindel\"of type, the asymptotic behaviour 
in both directions is related by analytic continuation. 

A slight generalization of the above virtual Compton amplitude with 
imaginary energy, which contains an additional mass variable $\mu$ for 
convergence, can be written as an integral over a matrix element in an 
effective Euclidean field theory. Using standard methods, this matrix 
element is expressed in terms of a Euclidean functional integral with a 
simple effective Lagrangian, which is given explicitly. The essential 
proposal of our paper is to attempt an evaluation of this functional 
integral using genuinely non-perturbative methods, e.g., lattice 
Monte-Carlo simulations. Finally, an analytic continuation to $\mu=0$ and 
to the real axis in the complex energy plane should allow the extraction 
of the desired asymptotic small-$\xbj$ limit of DIS structure functions. 

To illustrate how our approach works, we have explicitly performed all 
necessary steps in a simple model with free scalar fields. In this case, 
the Euclidean path integral is given by two tree-level Feynman diagrams and 
the subsequent analytic continuations in $\mu$ and the energy variable can 
be carried out explicitly. As expected, the final result for the structure 
function is in agreement with a direct diagrammatic calculation in 
Minkowski space. 

We have shown that in the free-field case the high-energy limit of the 
virtual Compton amplitude is governed by the behaviour of the effective 
Euclidean theory near the critical point $r=0$, where $r$ is our parameter 
defined in Eq.~(\ref{reta}). We have then assumed that also in the 
effective theory with interaction a critical point occurs at $r=0$ and 
discussed the possible consequences of this for the hard and soft pomeron 
regimes of the structure function. 

We do not claim that we have found the optimal method for an analytic 
continuation of high-energy amplitudes. In fact, in Sect.~\ref{alt} of this 
paper we introduce two alternative possibilities. They have the advantage 
that the Euclidean field theory obtained is based on the conventional 
Hamiltonian as opposed to the effective Hamiltonian required in the 
original approach. However, in both cases it appears to be more difficult 
to recover the small-$\xbj$ limit of structure functions from the final 
result of the calculation. 

A lot of work remains to be done before phenomenologically relevant 
information can be extracted from the approach suggested. Note first that 
all of the discussion in the present paper is based on a model with a 
scalar `photon' coupled to scalar partons. This has to be extended to the 
realistic case of a vector photon and QCD. However, we do not expect any 
fundamental problems with this generalization. 

Furthermore, the feasibility of a lattice calculation or any alternative 
non-per\-tur\-bative treatment of our Euclidean path integral expression has 
to be evaluated. Note that our effective Lagrangian contains an imaginary 
part, which, even though it does not raise any fundamental problems, may 
lead to technical difficulties on the lattice. Note also that the 
subsequent analytic continuation back to the physical region could prove 
highly non-trivial. 

On a more fundamental level, one has to ask whether better methods for the 
treatment of the small-$\xbj$ limit of DIS in a Euclidean theory exist. We 
have attempted to address this question in our cursory discussion of 
alternative methods of analytic continuation. However, we are not able to 
give a general answer at present. 

Given the above reservations, the importance of our results lies in their 
potential to relate small-$\xbj$ cross sections in DIS to quantities 
derivable from an effective Euclidean field theory. One can then hope to 
understand the limit $\xbj\to 0$ in terms of the critical behaviour in this 
Euclidean theory using all the tools which have been developed in 
statistical mechanics for the study of criticality. Also one can hope 
to calculate the relevant quantities on the lattice. This goal is 
fundamental since lattice calculations remain practically the only 
non-perturbative method in QFT that can be derived strictly from first 
principles. Thus, it is certainly worthwhile to continue the exploration of 
possibilities to obtain high-energy or small-$\xbj$ cross sections from 
Euclidean quantities.\\[.4cm]
{\bf Acknowledgements}\\[.1cm]
The authors are grateful to A. Donnachie, H.G. Dosch, P.V. Landshoff, 
F. Lenz and H.Ch. Pauli for useful discussions.

\section*{Appendix}
\begin{appendix}
\section{Considerations concerning the analytic continuation of amplitudes 
in the $\nu$ plane}\label{rac} 
\setcounter{equation}{0}
\renewcommand{\theequation}{\ref{rac}.\arabic{equation}}
Here we will first discuss the analytic continuation of $T_r(\nu,Q^2)$, 
Eq.~(\ref{trm}), from real values of $\nu$ into the half plane Im$\,\nu>0$, 
where we have
\be
\mbox{Im}\,\nu>\mbox{Im}\,\sqrt{\nu^2+Q^2}\ge 0\,.\label{die}
\ee
Proof: the second inequality is obvious. The first inequality holds for 
imaginary $\nu$. By continuity, the inequality can only be violated if 
there exists some $\nu$ in the upper half plane such that
\be
\mbox{Im}\,\nu=\mbox{Im}\,\sqrt{\nu^2+Q^2}\,.
\ee
Thus, for some $a\in I\!\!R$, 
\be
\sqrt{\nu^2+Q^2}=\nu+a\,.
\ee
Squaring this equation, one finds 
\be
Q^2=2\nu a+a^2\,,
\ee
which can not be fulfilled for non-zero $Q$ and Im$\,\nu>0$. Therefore, the 
first inequality in Eq.~(\ref{die}) holds everywhere in the upper half 
plane and our proof is complete. 

With Eq.~(\ref{die}) we find that the factor exp$(ix^0\nu-ix^3\sqrt{\nu^2+ 
Q^2})$ in Eq.~(\ref{trm}) decreases exponentially for Im$\,\nu>0$ as 
$x^0\to\infty$ with $|x^3|/x^0\le 1$. Thus the integral in Eq.~(\ref{trm}) 
should be well convergent and define an analytic function in $\nu$ for 
Im$\,\nu>0$. There is still the possibility of $T_r(\nu,Q^2)$ having a cut 
on the imaginary $\nu$ axis for $0\le\mbox{Im}\,\nu\le Q$ due to the 
explicit factor $\sqrt{\nu^2+Q^2}$ in Eq.~(\ref{trm}). But using 
Eq.~(\ref{rot}) it is easy to see that the values of $T_r(\nu,Q^2)$ when 
approaching the imaginary axis from both sides are equal. Then the ``edge of 
the wedge'' theorem (cf., e.g., \cite{eotw}) guarantees that $T_r(\nu,Q^2)$ is 
also analytic for $0<\mbox{Im}\,\nu<Q$. The point $\nu=iQ$ could still be 
an isolated singularity but this is incompatible with the square root in 
Eq.~(\ref{trm}). This concludes our discussion of the analyticity of 
$T_r(\nu,Q^2)$ for Im$\,\nu>0$. 

The discussion of the analyticity of $T_+(\nu,Q^2,\mu)$ defined as in 
Eq.~(\ref{tp}) but with $i\eta$ replaced by $\nu$ is analogous. The proof 
of Eq.~(\ref{trtp}) for $0<\eta<Q$ is straightforward using 
Eq.~(\ref{rot}). 

In the remainder of this appendix we will discuss the connection between 
the limits $\nu\to\infty$ either on the real or the positive imaginary 
axis. We will outline one possibility of specifying the requirements for 
guaranteeing that the analytic continuation of the asymptotic behaviour of 
the amplitude from the real to the imaginary axis is possible. 

The retarded amplitude $T_r(\nu,Q^2)$ is an analytic function of $\nu$ with 
cuts along the positive and negative real axis (Fig.~\ref{nup}). Let us 
assume that there exists a function $f(\nu)$ analytic for Im$\,\nu>0$ such 
that $f(\nu)\neq 0$ for Im$\,\nu\ge 0$, and 
\be
R(\nu)\equiv \frac{T_r(\nu,Q^2)}{f(\nu)}\label{rdef}
\ee
is a bounded function, $|R(\nu)|\le\,$const. for Im$\,\nu\ge 0$. (Since 
$Q^2$ is kept fixed we suppress this argument in $R(\nu)$ and $f(\nu)$.) 
Let us furthermore assume that 
\be
\lim_{\eta\to\pm\infty}R(\eta)=C_{\pm}\,.\label{cpm}
\ee
Then the Phragm\'{e}n-Lindel\"of theorem (see theorem 5.64 of~\cite{pl}) 
states that $C_+=C_-$ and that $R(\nu)$ approaches the same limit $C_+$ 
along any ray $\nu=\eta\exp(i\phi)$, $0\le\phi\le\pi$, $\phi=$\,const., 
$\eta\to +\infty$. Thus we have also on the imaginary axis 
\be
\lim_{\eta\to +\infty}R(i\eta)=C_+\,.\label{cpl}
\ee
Suppose now that from the study of $T_r(i\eta,Q^2)$, e.g., by lattice 
calculations, we can deduce the asymptotic behaviour for $\eta\to\infty$, 
construct a suitable function $f(i\eta)$ and deduce the value of the 
constant $C_+$ [Eq.~(\ref{cpl})] with $C_+\neq 0$. 

Typically one would try to fit lattice data to functions reflecting our 
general expectations concerning the high-energy behaviour of amplitudes 
in QFT like $f(\nu)=(\nu-\nu_0)^\alpha$ or $f(\nu)=(\nu-\nu_0)^\alpha 
\ln(\nu-\nu_0)$. Here $\alpha$ should be taken as a real constant and 
$\nu_0$ as a constant with Im$\,\nu_0<0$. With the assumptions specified in 
Eqs.~(\ref{rdef}) and (\ref{cpm}) we get for the physical amplitude on the 
real axis:
\be
T_r(\nu,Q^2)\,\longrightarrow\,C_+f(\nu)\qquad\mbox{for}\qquad\nu\to\infty
\,.
\ee
To summarize: if we {\it assume} that the limit of $T_r(\nu,Q^2)$ for 
$\nu\to\infty$, $0\le\mbox{arg}\,\nu\le\pi$ is governed by a suitable 
analytic function $f(\nu)$, then this function can be {\it determined} from 
the study of $T_r(\nu,Q^2)$ on the positive imaginary axis. 

We have only outlined here the simplest assumptions making the analytic 
continuation of the high-energy limits on the real and imaginary axes 
possible. No doubt, using the methods of~\cite{cm} one can relax these 
assumptions and still get useful relations between the two limits.

\section{The functional integrals for ${\cal M}_E$ and ${\cal M}_M$}
\label{fic}
\setcounter{equation}{0}
\renewcommand{\theequation}{\ref{fic}.\arabic{equation}}
In this appendix we give the details of the derivation of Eq.~(\ref{emer}). 
We start with Eq.~(\ref{eme}) and using Eq.~(\ref{lsze}) rewrite it as 
\bea
{\cal M}_E(\tau,r)&\!\!\!\!\!=\!\!\!\!&\frac{1}{Z}\lim_{\tau_i\to-\infty
\atop\tau_f\to+\infty}\lim_{\tau_i'\to-\infty\atop\tau_f'\to+\infty}\,
e^{(\tau_f-\tau_i)M}\,\langle 0|\,e^{-(\tau_f'-\tau_f)H_{eff}(r)}\,A(0)\,
e^{-(\tau_f-\tau)H_{eff}(r)}\,\label{mea}\\
\nonumber\\
&&\times J(0)\,e^{-(\tau-0)H_{eff}(r)}\,J(0)\,e^{-(0-\tau_i)H_{eff}(r)}\,
A^\dagger(0)\,e^{-(\tau_i-\tau_i')H_{eff}(r)}\,|0\rangle\Bigg/
e^{-(\tau_f'-\tau_i')E_0}\,.\nonumber
\eea
Here we have introduced further times $\tau_i'$, $\tau_f'$ with $\tau_i'
<\tau_i<\tau_f<\tau_f'$ and the limits in Eq.~(\ref{mea}) are to be taken 
in the order indicated there. 

Let $|\phi\rangle$ be eigenstates of the field operator $\Phi(x)$ at
$x^0=0$:
\be
\Phi(\vec{x},0)|\phi\rangle=\phi(\vec{x})|\phi\rangle\,,
\ee
where $\phi(\vec{x})$ are classical functions. We have 
\be
\int D\phi\,|\phi\rangle\langle\phi|={\bf 1}\,.\label{one}
\ee
The basic relation of the path integral formalism~\cite{pi} 
discussed in many textbooks (see, e.g.,~\cite{wein}) is in our case 
\bea
\langle \phi^{(2)}|\,e^{-\Delta\tau\,H_{eff}(r)}\,|\phi^{(1)}\rangle &=&
\int D\pi^{(2)}\,e^{i\int d^3\vec{x}\left[\pi^{(2)}(\vec{x})\left(\phi^{(2)}
(\vec{x})-\phi^{(1)}(\vec{x})\right)+i\Delta\tau\,{\cal H}_r\left(\pi^{(2)}
(\vec{x}),\phi^{(2)}(\vec{x})\right)\right]}\nonumber
\\
&&+{\cal O}(\Delta\tau^2)\,,\label{pif}
\eea
where the measure includes a factor $1/(2\pi)$ for each $\pi^{(2)}(\vec{x} 
)$ integration. Here $\pi^{(2)}(\vec{x})$ is the classical momentum field 
and ${\cal H}_r$ the classical Hamilton density corresponding to $H_{eff} 
(r)$:
\be
{\cal H}_r(\pi,\phi)=\frac{1}{2}\pi(\vec{x})^2+(1-r)\pi(\vec{x})
\partial_3\phi(\vec{x})+\frac{1}{2}\left(\vec{\nabla}\phi(\vec{x})\right)^2
+\frac{1}{2}m^2\phi(\vec{x})^2+\frac{\lambda}{4!}\phi(\vec{x})^4\,.
\ee
Next we choose a grid on the time interval $[\tau_i',\tau_f']$: $\tau_i' 
\equiv\tau^{(0)}<\tau^{(1)}<\tau^{(2)}\cdots<\tau^{(N)}\equiv\tau_f'$, 
where also $\tau_i$, 0, $\tau$, $\tau_f$ should occur as intermediate 
points $\tau^{(j)}$. We discuss the factor exp$[-(\tau_f'-\tau_i')E_0]$ 
which we rewrite using Eqs.~(\ref{one}) and (\ref{pif}) as:
\bea
e^{-(\tau_f'-\tau_i')E_0}&=&\langle 0|\,e^{-(\tau_f'-\tau_i')H_{eff}(r)}|0 
\rangle
\\
&=&\langle 0|e^{-(\tau^{(N)}-\tau^{(N-1)})H_{eff}(r)}\,\cdots\,
e^{-(\tau^{(1)}-\tau^{(0)})H_{eff}(r)}\,|0\rangle\nonumber
\\
&=&\int\prod_{j=0}^N D\phi^{(j)}\,\langle 0|\phi^{(N)}\rangle\langle
\phi^{(N)}|\,e^{-(\tau^{(N)}-\tau^{(N-1)})H_{eff}(r)}\,|\phi^{(N-1)}
\rangle\nonumber
\\
&&\cdots\langle\phi^{(1)}|\,e^{-(\tau^{(1)}-\tau^{(0)})H_{eff}(r)}\,|
\phi^{(0)}\rangle\langle\phi^{(0)}|0\rangle\,,\nonumber
\eea
\bea
e^{-(\tau_f'-\tau_i')E_0}&=&\lim_{N\to\infty\atop\Delta\tau\to 0}
\int\prod_{j=0}^N D\phi^{(j)}\prod_{j=1}^N D\pi^{(j)}\langle 0|\phi^{(N)}
\rangle\label{efi}
\\
&&\hspace*{-3cm}
\times\prod_{j=1}^N e^{i\int d^3\vec{x}^{(j)}\left[\pi(\vec{x}^{(j)},\tau^{(j)})
\left(\phi(\vec{x}^{(j)},\tau^{(j)})-\phi(\vec{x}^{(j)},\tau^{(j-1)})\right)
+i(\tau^{(j)}-\tau^{(j-1)}){\cal H}_r(\pi(\vec{x}^{(j)},\tau^{(j)}),
\phi(\vec{x}^{(j)},\tau^{(j)})\right]}\langle\phi^{(0)}|0\rangle\,.
\nonumber
\eea
Here we set $\phi^{(j)}(\vec{x})\equiv\phi(\vec{x},\tau^{(j)})$ and 
$\pi^{(j)}(\vec{x})\equiv\pi(\vec{x},\tau^{(j)})$. 

Now the integration over the $\pi$ fields in Eq.~(\ref{efi}) can be 
performed since the integral is of Gaussian type. In the limit 
$N\to\infty$, $\Delta\tau\to 0$ we get
\be
e^{-(\tau_f'-\tau_i')E_0}={\cal C}\int D\phi\,e^{-\int_{\tau_i'}^{\tau_f'}
d\tau\int d^3\vec{x}\,{\cal L}_{E,\,eff}(\vec{x},\tau)}\,,
\ee
where ${\cal L}_{E,\,eff}$ is given in Eq.~(\ref{el}) and ${\cal C}$ is a 
constant (which is infinite). Performing the same steps for the matrix 
element in the numerator in Eq.~(\ref{mea}) and taking the limit 
$\tau_i'\to -\infty$, $\tau_f'\to+\infty$ we get the expression for 
${\cal M}_E(\tau,r)$ of Eq.~(\ref{emer}). If we perform the analogous 
steps for the Minkowskian matrix element, Eq.~(\ref{mme}), we get: 
\be
{\cal M}_M(t,r)=\frac{1}{Z}\lim_{t_i\to-\infty\atop t_f\to+\infty}
\,e^{i(t_f-t_i)M}\,{\cal Z}^{-1}\int D\phi\, a(t_f)j(t)j(0)a^\dagger
(t_i)\,e^{i\int d^4x\,{\cal L}_{M,\,eff}}\,,
\ee
\be
{\cal Z}=\int D\phi\,e^{i\int d^4x\,{\cal L}_{M,\,eff}}\,,
\ee
with
\be
{\cal L}_{M,\,eff}=\frac{1}{2}\Big(\partial_0{\phi}(x)-(1-r)\,\partial_3
\phi(x)\Big)^2-\frac{1}{2}\Big(\nabla\phi(x)\Big)^2-\frac{m^2}{2}\phi(x)^2-
\frac{\lambda}{4!}\phi(x)^4\,
\ee
and
\bea
a(t_f)&=&\int_{x^0=t_f}d^3\vec{x}\,\Big(i\partial_0{\phi}
(x)-i(1-r)\,\partial_3\phi(x)+M\phi(x)\Big)\,,\\
a^\dagger(t_i)&=&\int_{x^0=t_i}d^3\vec{x}\,\Big(-i\partial_0{\phi}
(x)+i(1-r)\,\partial_3\phi(x)+M\phi(x)\Big)\,.
\eea
\end{appendix}

\end{document}